\documentclass[aps,prd,a4paper,twocolumn,showpacs,superscriptaddress,preprintnumbers, floatfix,nofootinbib]{revtex4-1}
\usepackage{amsmath}
\usepackage{hyperref, url}
\usepackage{graphicx,subfigure}
\usepackage[normalem]{ulem}
\usepackage{epsfig}
\usepackage{color}
\usepackage{multirow}
\usepackage{array}
\usepackage[utf8]{inputenc}
\usepackage[T1]{fontenc}
\usepackage[dvipsnames]{xcolor}
\usepackage{comment}
\hypersetup{colorlinks   = true,
            urlcolor     = blue,
            citecolor    = blue,
            linkcolor    = blue,
            menucolor    = blue,
            anchorcolor  = blue,
            filecolor    = blue}

\enlargethispage{\baselineskip}

\usepackage[utf8]{inputenc}  
\usepackage[T1]{fontenc}

\usepackage{graphicx}

\def\a{\alpha}     
    
 \def\k{\kappa}  \def\m{\mu} \def\n{\nu}

 \newcommand{\Lcal}{{\mathcal L}}

\newcommand{\approptoinn}[2]{\mathrel{\vcenter{
  \offinterlineskip\halign{\hfil$##$\cr
    #1\propto\cr\noalign{\kern2pt}#1\sim\cr\noalign{\kern-2pt}}}}}

\emergencystretch=\maxdimen
\usepackage{bm,epstopdf,natbib,hyperref,color,verbatim,multirow,bm,tikz,xcolor}
\hypersetup{colorlinks=true,urlcolor=blue,citecolor=blue,linkcolor=blue,menucolor=blue,anchorcolor=blue,filecolor=blue}
\everymath{\displaystyle}
\definecolor{lime}{HTML}{A6CE39}
\DeclareRobustCommand{\orcidicon}{
	\begin{tikzpicture}
	\draw[lime, fill=lime] (0,0) 
	circle [radius=0.16] 
	node[white] {{\fontfamily{qag}\selectfont \tiny ID}};
	\draw[white, fill=white] (-0.0625,0.095) 
	circle [radius=0.007];
	\end{tikzpicture}
	\hspace{-3mm}
}
\foreach \x in {A, ..., Z}{\expandafter\xdef\csname orcid\x\endcsname{\noexpand\href{https://orcid.org/\csname orcidauthor\x\endcsname}
			{\noexpand\orcidicon}}
}

\begin{document}

\title{QUPITER – Space Quantum Sensors for Jovian-Bound 
Dark Matter}

\author{Yu-Dai Tsai\hspace{-1mm}\orcidA{}}
\email{yt444@cornell.edu; yudaitsai.academic@gmail.com}
\affiliation{Los Alamos National Laboratory (LANL), Los Alamos, NM 87545, USA}

\author{Fazlollah Hajkarim\hspace{-1mm}\orcidB{}}
\email{fazlollah.hajkarim@ou.edu}
\affiliation{Department of Physics and Astronomy, University of Oklahoma, Norman, OK 73019, USA}

\date{\today}

\begin{abstract}
\noindent 

We propose utilizing space quantum sensors to detect ultralight dark matter (ULDM) bound to planetary bodies, focusing on Jupiter as the heaviest planet in the solar system. Leveraging Jupiter's deep gravitational potential and the wealth of experience from numerous successful missions, we present strong sensitivity projections on the mass and couplings of scalar ULDM. Future space missions offer unique opportunities to probe the ULDM interactions using quantum sensors, including superconducting quantum interference device (SQUID) magnetometers. By measuring dark matter-induced magnetic field oscillations, we expect to achieve sensitivity orders of magnitude beyond the terrestrial probes and significantly improve detection prospects of theoretically motivated ULDM candidates.

\end{abstract}

\maketitle

\section{Introduction}

Dark matter (DM) is the most significant remaining puzzle to complete our understanding of particle physics and cosmology. 
The pursuit of DM has motivated theoretical and experimental efforts to illuminate its characteristics.
One of the well-motivated classes of DM models is 
ultralight dark matter (ULDM), which could be QCD axions or a more generic scalar/vector dark matter. Experimental tools like atomic clocks and precision spectroscopy have provided the possibility of probing the nature of ULDM and its possible interactions at different frequencies~\cite{Feng:2010gw,Budker:2013hfa,Graham:2015ifn,Antypas:2020rtg,Antypas:2019qji, Gramolin:2020ict,Blanco:2023qgi,Antypas:2020rtg,Oswald:2021vtc,Antypas:2022asj,Tretiak:2022ndx,Bloch:2023wfz,Bloch:2023uis,Adams:2022pbo}, and point us to the potential discovery of ULDM with precision measurements. Astrophysical and cosmological observations have have set curcial limits on ULDMs~\cite{Poddar:2019zoe,Tsai:2023zza,Bar:2019bqz,Tsai:2021lly,Antypas:2022asj,Tsai:2020vpi,JacksonKimball:2023ers,Kennedy:2020bac,Sibiryakov:2020eir,Hui:2016ltb,Baryakhtar:2017ngi,Hlozek:2014lca,Davoudiasl:2019nlo,Roy:2019esk,Vagnozzi:2022moj,Gau:2023rct,Shakeri:2022usk,Fortin:2018aom,Green:2022hhj,Baryakhtar:2022hbu}.

Furthermore, detection methods like space-based quantum sensors have been proposed to probe different types of DM, especially when they are bound gravitationally to the Sun.  This will push the observational limits beyond Earth-based experiments \cite{Tsai:2021lly}. 
The observation of variation of Jupiter’s magnetic field has given us some constraints on DM candidates like dark photons. This will improve the exploration of the DM parameter space, especially at low masses \cite{Yan:2023kdg}. 
These efforts have improved our understanding of ultralight dark matter properties and guided us in the development of next-generation experiments to search for these elusive particles  \cite{Banerjee:2019epw,Hees:2018fpg}. In addition, using asteroidal and planetary motions general dark matter overdensity can be constrained~\cite{pitjev2013constraints,Tsai:2022jnv} and ULDM couplings~\cite{Tsai:2021irw,Tsai:2023zza}.

In particular, Jupiter's environment provides interesting opportunities for studying physics beyond the Standard Model (SM), including DM and its mediators to SM interactions (e.g., dark photons). 
Magnetic field measurements from the JUNO spacecraft have set limits on dark photon masses and kinetic mixing to SM photons, independent of their role as DM~\cite{Yan:2023kdg}.

In this paper, we study the effect of ULDMs around Jupiter and the possibility of probing it using a sensitive magnetometer. 
Advanced quantum sensors, such as atomic clocks and magnetometers, are used by missions like JUNO, JUICE, Europa Clipper, and Europa Lander. They will also be exploited by the upcoming Io Volcano Observer, which can monitor magnetic oscillations and gravitational signatures to detect ultralight dark matter, accessing frequency ranges unattainable by terrestrial instruments. Additionally, missions investigating Jupiter’s subsurface oceans, including Europa Clipper and the Europa Lander, can help explore ULDMs bound to Jupiter as well. Including DM probes into these ongoing and future missions significantly advances ultralight dark matter research within our solar system through innovative space-based technologies. We study the density profile of dark matter around planets and focus on Jupiter as a prime example, and discuss the experimental setup using magnetometers to detect the scalar ULDM. 

The paper is organized as follows: In section~\ref{sec:jup-den}, we consider the overdensity of DM bound to Jupiter. In section~\ref{sec:QS}, we discuss the magnetometer and shielding required to study ULDM effects, and we conclude in section~\ref{sec:conclusion}.

 \section{Jupiter Dark Matter Overdensity}
\label{sec:jup-den}

The average DM density is about $0.4 \, \text{GeV/cm}^3$ \cite{Bertone:2004pz,Cirelli:2024ssz} in the local galactic environment. However, there might be fluctuations in the density of DM, especially at the scale of the solar system.
We currently have an upper limit on the density of DM that is obtained by the planetary and asteroidal motions~\cite{pitjev2013constraints,Tsai:2022jnv}. Also, simulations predict overdensity for some models and formation scenarios~\cite{Bar:2019bqz,Banerjee:2019xuy, Banerjee:2019epw,Banerjee:2019epw,Banerjee:2019xuy,Budker:2023sex}. This makes it crucial to examine the density and presence of DM with dedicated missions.

A compelling possibility is that DM may become gravitationally bound to objects within the solar system, such as the Sun or Earth, resulting in modifications to the local DM density. Previous works and simulations suggest that ULDM can potentially form a ``subhalo" structure, stabilized by the celestial object's gravitational potential. 
The maximally allowed density profile of such a subhalo for a ULDM mass $m_\phi$ is given by
\cite{Banerjee:2019epw}

\begin{equation}
    \rho_\star \propto 
    \begin{cases}
        \displaystyle{\exp\left(-\frac{2r}{R_\star}\right)} & \text{for } R_\star > R_{\rm ext}, \\
        \displaystyle{\exp\left(-\frac{r^2}{R_\star^2}\right)} & \text{for } R_\star \leq R_{\rm ext}.
    \end{cases}
\end{equation}
$R_\star$, the characteristic radius of the halo for $M_\star \ll M_{\rm ext}$, is determined by 
\begin{equation} \label{eq:R_star}
R_\star \equiv 
\begin{cases}
    \displaystyle{\frac{M_{\text{Pl}}^2}{m_\phi^2} \frac{1}{M_{\rm ext}}} & \text{for } R_\star > R_{\rm ext}, \\
    \displaystyle{\left( \frac{M_{\text{Pl}}^2}{m_\phi^2} \frac{R_{\rm ext}^3}{M_{\rm ext}} \right)^{1/4}} & \text{for } R_\star \leq R_{\rm ext},
\end{cases}
\end{equation}
where $R_\star$ depends on the mass of the provider of the external potential $M_{\rm ext}$, and the ULDM mass $m_\phi$. $M_{\text{Pl}}$ is the Planck mass, $M_{\text{Pl}}=1.22\times10^{19}$ GeV, and $R_{\rm ext}$ is the radius of the external object. 
$M_{\rm ext}=M_{\rm Jupiter}$ would be the mass of the external host body, in our consideration, a mission near Jupiter.
The coherence time $\tau_\star$ of the scalar subhalo oscillations is crucial for the sensitivity of direct-detection experiments searching for ULDM and is defined as~\cite{Antypas:2019qji,Banerjee:2019epw,Tsai:2021lly}
\begin{equation}
    \tau_\star =\frac{1}{m_{\phi}\beta^2} = m_\phi R_\star^2,
\end{equation}
The coherence time is typically longer than $10^3-10^4$ seconds.

One can derive the upper bounds on the overdensities of these subhalo structures by studying the precision data of the orbits around the host body. The subhalo around the Sun, i.e., the solar halo, can be constrained by planetary ephemerides~\cite{Pitjev:2013sfa}, and more recently, dedicated asteroidal studies~\cite{Tsai:2022jnv,Tsai:2023zza}. The Earth-bound halo can be constrained by Lunar Laser Ranging (LLR)~\cite{Adler:2008rq} and further explored with quantum sensors in~\cite{Antypas:2019qji,Banerjee:2019epw,Tsai:2021lly}. 
In addition, given that the DM subhalo size depends on the mass of the celestial body, $M_{\rm ext}$. A dedicated DM probe to planets like Jupiter would be important for covering ULDM parameters not covered by similar terrestrial or solar probes. When considering the overdensity of dark matter bound to Jupiter, we impose the following condition:
\begin{equation} \label{eq:M_star_constraint}
    M_\star < {\rm Min}
    \left
    [\frac{1}{2} M_{\rm Jup} \left(\frac{R_\star}{R_{\rm Jup}}\right)^3, 1\% M_{\rm Jup}\right].
\end{equation}
This condition ensures that the ULDM mass does not exceed 1 percent of Jupiter's mass; in the situation where the subhalo radius is smaller than that of Jupiter, it does not replace the Jupiter core with a ULDM core.

\subsection{Jupiter and Jovian Moon Missions}
\label{sec:jup-mission}

Table~\ref{table:jupiter_moons} represents information about Jupiter's moons, including the orbital period, semi-major axis, and eccentricity of each moon. 
Also, each space mission that can probe and spend some time around each moon is listed in Table~\ref{table:jupiter_moons}. 
In Table~\ref{tab:jupiter_missions}, we have briefly shown the missions around Jupiter and its moons and the time of each mission around each moon. These tables are general information for the reader and include future planned or proposed missions that can be used as a potential carrier of the experimental setup, similar to what is mentioned in this study. More details of these missions are presented in the Appendix~\ref{app:jupmis}.

\begin{table*}[!]
\centering
\begingroup
\setlength{\extrarowheight}{1.5 mm}

\begin{tabular}{|>{\centering\arraybackslash}m{1.5cm}|>{\centering\arraybackslash}m{6.5cm}|>{\centering\arraybackslash}m{3cm}|>{\centering\arraybackslash}m{3.5cm}|>{\centering\arraybackslash}m{2cm}|}
\hline
\textbf{Moons} & \textbf{Missions} & \textbf{Orbital Period} & \textbf{Semi-major Axis} & \textbf{Eccentricity} \\
\textbf{} & \textbf{} & \textbf{(days)} & \textbf{(km/AU)} & \textbf{} \\
\hline
Io & JUNO / JUICE / IVO & 1.77 & 421,700 / 2.82$\times 10^{-3}$ & 0.0041 \\ 
\hline
Europa & JUICE / Europa Clipper / Europa Lander & 3.55 & 671,034 / 4.49$\times 10^{-3}$ & 0.0094 \\
\hline
Ganymede & JUNO / JUICE / Europa Clipper & 7.15 & 1,070,412 / 7.16$\times 10^{-3}$ & 0.0013 \\
\hline
Callisto & JUNO / JUICE & 16.69 & 1,882,700 / 1.26$\times 10^{-2}$ & 0.0074 \\
\hline
\end{tabular}

\endgroup
\caption{
Characteristics of a selection of Jupiter's Moons~\cite{rogers1995giant,bagenal2006jupiter,bland2010effects,bagenal2007jupiter,carr1998evidence, kivelson2000galileo, grasset2013jupiter, 2009euro.book.....P, schenk2008true, greenberg2005europa}
 and the relevant missions around them \cite{witasse2021juice, grasset2013review, dougherty2012jupiter, bolton2017jupiter, matousek2007JUNO, connerney2017jupiter, adriani2018clusters, mcewen2014io, mcewen2014io, mcewen2014io, williams2011volcanism, phillips2014europa, pappalardo2013science, pappalardo2024science, vance2023investigating, pappalardo2024science,hamilton2024comparing,bhardwaj2002i0,de2024isotopic,davies2024io,tyler2015tidal}. 
}
\label{table:jupiter_moons}
\end{table*}

\section{Ultralight Dark Matter Signatures}
\label{sec:QS}

We aim to probe the electron and photon couplings to ULDM $\phi$, through the following interaction Lagrangian~\cite{DiLuzio:2020wdo}
\begin{eqnarray} 
  \label{eq:Lag}
 \Lcal 
 &\supset&\phi \left(g_e \bar{e}e+\frac{1}{4}g_\gamma F_{\mu\nu}F^{\mu\nu} \right)
 \\  
 &=&  \k \phi \left(d_{m_e}m_e\bar{e} e 
 + \frac{d_\a}{4}F_{\m\n}F^{\m\n}
 \right)
 .
\end{eqnarray} 
Here, $e$ denotes the electron field, $F^{\m\n}$ represents the electromagnetic field strength, and $\k = \sqrt{4\pi}/M_{\text{Pl}}$. Parameters $g_{e(\gamma)}$ are the coupling of $\phi$ to electrons (photons), with $g_\gamma$ with the dimension of the inverse energy scale, originating from a cut-off scale $\Lambda$ from the fundamental theory. $d_{m_e, \alpha}$ are dimensionless couplings according to the conventions of atomic, molecular, and optical (AMO) physics. A range of sensors have been applied to study ultralight dark matter, including precision clocks~\cite{Sherrill:2023zah} and resonant-mass detectors~\cite{Arvanitaki:2015iga}.In addition to precision magnetometers, they can also be applied to our proposal to probe dark matter overdensity around Jupiter.

The ULDM field $\phi$ oscillates as
\begin{eqnarray}
\phi = \phi_0 \cos(\omega t)\,,~~ \phi_0 = \frac{\sqrt{2 \rho_{\text{DM}}}}{m_\phi}\,.
\end{eqnarray}
We operate in a quasi-static configuration such that the ULDM wavelength ($\gtrsim$ 3\,km) is much longer than the experiment, and the integration time is much longer than  1/$\omega$, so we neglect the location-dependent phase~\cite{Tsai:2021lly,Bloch:2023uis}.

The oscillation of the ULDM field, when coupled to electrons, will induce a change in the electron mass and the Bohr magneton, as
\begin{eqnarray}
\frac{\delta \mu_B}{\mu_B} = -\frac{\delta m_e}{m_e} = -\frac{g_e}{m_e} \phi_0 \cos(\omega t)\,,
\end{eqnarray}
and induce an oscillation of the magnetization,
\begin{eqnarray}
\frac{\delta M}{M_0} = \frac{\delta \mu_B}{\mu_B} = -\frac{g_e}{m_e} \phi_0 \cos(\omega t)\,,
\end{eqnarray}
where the unperturbed permanent magnetization quantity is denoted by $M_0$, and $\delta B = 4\pi \delta M$. Furthermore, marginalizing the complicated geometric factors discussed in~\cite{Bloch:2023uis}, the photon coupling of the ULDM field also induces a magnetic field oscillation. The magnetic field oscillation of a permanent magnet, induced by the time-varying ULDM field, can be measured via a magnetometer with the resulting magnetic signal as:
\begin{eqnarray}
\delta B = B_0 g \phi_0 \cos(\omega t), \label{eq:magnetic_field}
\end{eqnarray}
where $g\equiv-g_e/m_e+g_\gamma$, $B_0 = 4\pi M_0$ is the unperturbed magnetic field along a cylindrical magnet, and $\omega=2\pi f$.

We consider the sensitivity projection derived from the signal-to-noise ratio (SNR) of the observation. The dispersion velocity of the sub-halo, $v$, and the coherent time, $\tau_{\star}$, are important in the estimation. $\tau_{\star}=(\xi f)^{-1}$ is the minimal stable timescale for the DM configuration \cite{Tsai:2021lly}. $\xi = \beta^2$, which depends on the sub-halos, and $f$ is the frequency of oscillation of a scalar field with mass $m_{\phi}$.
For the  integration time $t > (\xi f)^{-1}$, the signal-to-noise ratio is obtained in \cite{Bloch:2023uis,Budker:2013hfa}
\begin{eqnarray}
\text{SNR} = \mathcal{B} ~ \frac{g\sqrt{\rho_{\text{DM}}}}{m_{\phi}} \left(\frac{t}{\xi f}\right)^{1/4}\, \label{eq:snr_1}
\end{eqnarray}
where $\mathcal{B}\equiv B_0 /\sqrt{S_{B}}$. Using this signal-to-noise ratio at order one, we can obtain projected sensitivity curves. 

A state-of-the-art magnetic field sensitivity using superconducting quantum interference device (SQUID) technology was achieved by the SHAFT experiment~\cite{Gramolin:2020ict}, with the noise performance parameters of $\sqrt{S_B}\simeq150 {\rm ~aT ~Hz}^{-\frac12}$. 
While this performance is degraded at frequencies below $\mathcal{O}(10\mathrm{kHz})$, we will consider a configuration that achieves a similar noise performance as a benchmark for our sensitivity projection.

\subsection{Proposed Detector Configuration}

Here, we consider a probe within the frequency range of $\sim 1$ Hz to $100$ kHz (mass range from $\sim 10^{-14}$ to $10^{-8}$ eV) by measuring the magnetic field oscillation induced in a permanent magnet by ULDM field oscillation~\cite{Bloch:2023uis}. 
More specifically, we use the SQUID magnetometer as an example to conduct the measurement, as it is one of the most sensitive instruments for detecting magnetic fields. SQUIDs are superconducting devices, and they must be kept below their critical temperature $T_c$. Cryogenics may be needed and is further discussed in Sec.~\ref{sec:bkg_shielding}.

\begin{figure}[t]
 \includegraphics[scale=0.27]{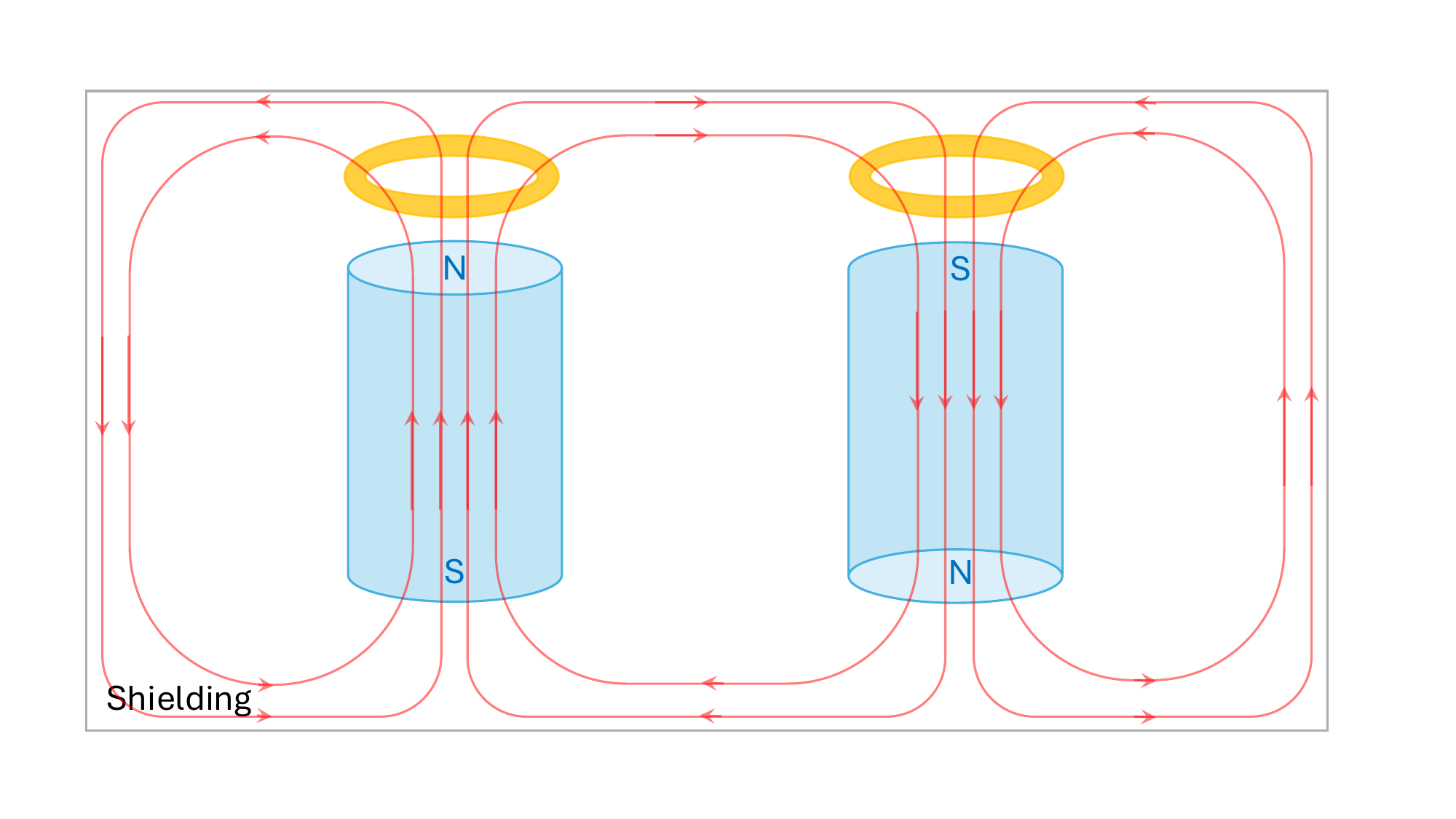}
 \caption{
Conceptual illustration of the proposed experimental setup. The yellow loops denote SQUID magnetometer pickup loops installed at two locations to detect magnetic-field oscillations (see the main text for detailed discussions). The blue cylindrical objects represent permanent magnets, which serve as sources of magnetic fields, depicted by red arrows. To suppress external magnetic noise, such as that from Jupiter’s magnetosphere, the entire apparatus can be enclosed within a (superconducting) shield. 
 }
 \label{fig:exp}
\end{figure}
One possibility is to use a pair of permanent magnets arranged with opposite polarity, read out with a pair of SQUID magnetometers, illustrated in Fig.~\ref{fig:exp}. This configuration allows for common-mode rejection between the two channels, mitigating magnetic field noise, that may result from variations in Jupiter's field or from radio-frequency interference. If further shielding is required, the magnets may be enclosed in a superconducting shield~\cite{Fosbinder-Elkins:2017osp}.    
We will use SQUID detectors of a similar capability as in~\cite{Gramolin:2020ict} as a demonstration of our derivation.

\subsection{Background and Shielding}
\label{sec:bkg_shielding}

Precisely measuring the oscillation of the magnetic field on a spacecraft near Jupiter presents intriguing technical challenges and opportunities. Jupiter has a strong magnetic field environment, often referred to as its magnetosphere, which would be a background to our measurements. 
Although this depends on the detailed orbit for our experimental setup, we comment on the shielding and the benefit of the low temperature of the mission environment to help achieve this shielding.

Jupiter is a gaseous planet, and the magnetic field on the surface is around a few Gauss \cite{bagenal2007jupiter,connerney2018new,bolton2017jupiter}.
The environment of Jupiter and its impacts on the study of fundamental new physics have been discussed in \cite{Li:2022wix,Yan:2023kdg}, especially in terms of the magnetic field. Using the data obtained from the JUNO spacecraft, it is determined that the magnetic field is of the magnitude $\sim$ 0.1 Gauss ($10^{-5}$ T), with a variation at the nT-level with a $\sim$ 15-second period (see Fig. 2 of~\cite{Yan:2023kdg}). This is far below our frequency range of interest. However, the background noise in the target frequency range in our paper (10 to $10^5$ Hz) requires further investigation.

In general, there are two strategies regarding the magnetic field background. 
(1) Provide significant shielding for the Jupiter magnetic field, install an artificial permanent magnetic field inside the shielding, and measure the variation of the magnetic field provided by the artificial permanent magnetic.
(2) Directly study the oscillation of the Jupiter magnetic field induced by ULDM. 
In our study, we will consider strategy (1). Strategy (2) requires a dedicated analysis of the existing data, which we leave for future consideration.

A superconducting shield should work below its critical temperature ($T_c$). There are low-$T_c$ superconductors, such as niobium ($T_c \sim 9.2 \,\text{K}$) or lead ($T_c \sim 7.2 \,\text{K}$). These typically require cryogenic cooling with liquid helium ($T \sim 4.2 \,\text{K}$), making them ideal for high-precision applications like magnetometers~\cite{cardwell2003handbook,mele2019superconductivity,buckel2008superconductivity}. Other types can operate at high-$T_c$, like YBCO ($T_c \sim 90 \,\text{K}$), working at liquid nitrogen temperatures ($T \sim 77 \,\text{K}$). This will reduce cooling costs for moderate shielding needs. For ultra-sensitive quantum experiments and superconducting qubits, there are materials with extremely low $T_c$, such as aluminum ($T_c \approx 1.2 \,\text{K}$). They can be operational in the millikelvin range using dilution refrigerators ($T \sim 10^{-3} \,\text{K}$) \cite{cardwell2003handbook,mele2019superconductivity,buckel2008superconductivity}. The aforementioned setups can provide appropriate magnetic-field shielding and stability for cutting-edge applications such as quantum computing, precision measurements, and space-based detectors.

To reach the operational temperature limit for the magnetometers or other experimental devices in space, a cooling setup is required. This has been done in previous space missions. The cooling systems of the Planck, LiteBIRD, and James Webb Space Telescope (JWST) are essential for reducing thermal noise in their observations, especially in the infrared and microwave ranges. The Planck satellite was launched in 2009 to observe the Cosmic Microwave Background (CMB). It used passive and active cooling systems, including hydrogen sorption and dilution refrigerators. This allows the achievement of temperatures as low as 0.1 K for its High-Frequency Instrument (HFI) \cite{lamarre2003planck,aghanim2020planck,Morgante:2009dqw}. Similarly, LiteBIRD is another advanced upcoming satellite mission. It is designed to observe and measure the B-mode polarization patterns in the cosmic microwave background, aiming to probe the early universe. It will use radiative cooling along with mechanical refrigerators and advanced cryogenic systems, such as helium-based coolers and sub-Kelvin instruments, to reach temperatures between 0.1--0.3 K \cite{LiteBIRD:2022cnt,Matsumura:2013aja,Suzuki:2018cuy,LiteBIRD:2020zfx}. The JWST mission, which was launched in 2021, can probe infrared frequencies. It relies on passive cooling through a large sunshield and an active cryocooler for its Mid-Infrared Instrument (MIRI). This helps to achieve temperatures as low as 6.7 K, with the rest of the telescope operating around 40 K \cite{lightsey2012james,Gardner:2006ky,ressler2015mid}. These advanced cooling technologies are important to obtain the sensitivity required to detect faint astronomical signals. In our proposed experimental setup, the possible cooling of magnets to the required temperature around or below a few Kelvin can be achieved by the combination of cooling methods done in the previous space missions we discussed earlier. It should be noted that mechanical cryocoolers will induce mechanical vibrations which may limit the sensitivity at frequencies below $\mathcal{O}(10\,\mathrm{kHz})$.

If this experiment were to take place on a JUNO-like spacecraft, the magnetic field strength could be lower than what is measured at Earth's surface~\cite{Li:2022wix,Yan:2023kdg}. While further study is required to understand the background magnetic field in the specific frequency range associated with DM oscillations, the level of shielding provided on Earth may be enough for this experiment to operate. Moreover, it might be possible to use a combination of magnetic shielding methods to reduce the magnetic noise due to Jupiter and other possible backgrounds \cite{ma2020magnetic,gao2024analysis,volegov2004simultaneous,romalis2011atomic,fang2022high}. The shielding, in principle, can also have effects on the signals of dark matter, which are discussed in~\cite{Bloch:2023uis,Chaudhuri:2014dla,JacksonKimball:2016wzv}, and it is determined that the effect is insignificant for the scalar dark matter cases. 

\begin{figure*}[t]
 \centering
   \includegraphics[scale=0.85]{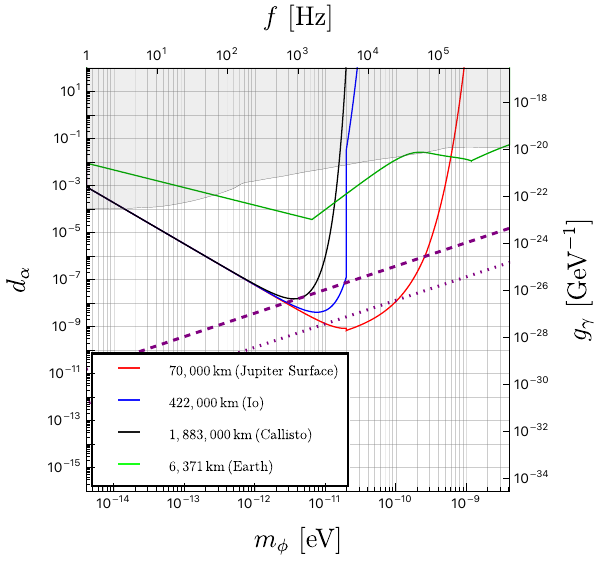}
   \includegraphics[scale=0.85]{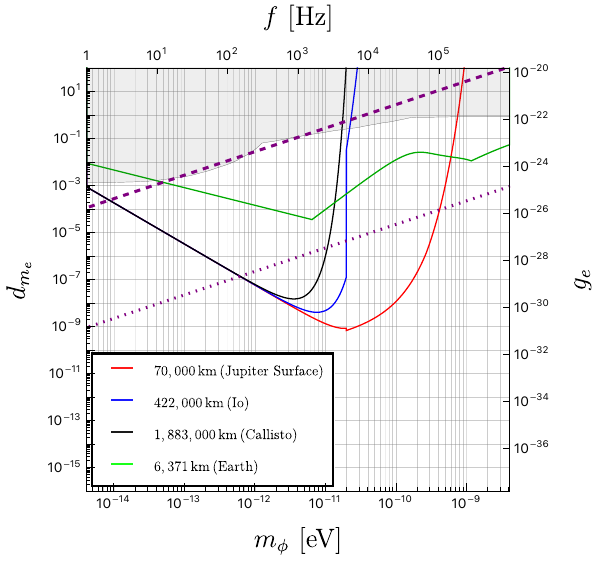}
   \caption{
     Sensitivity projections for the couplings of the scalar ULDM $\phi$ of mass $m_\phi$ to the photon, $d_\alpha$ (left panel), and to the electron, $d_{m_e}$ (right panel). The dotted and dashed purple lines represent theoretical predictions (see main text) \cite{Flacke:2016szy,Choi:2016luu,Tsai:2021lly}. The shaded gray region denotes constraints from equivalence principle experiments \cite{Wagner:2012ui,Berge:2017ovy,Hees:2018fpg}. The projections are shown for detectors placed on Earth, on the surface of Jupiter, and near the Jovian moons Io and Callisto. 
   }
   \label{fig:dem}
\end{figure*}

\subsection{Sensitivity Projections}

We present a sensitivity projection assuming $B_0 = 0.35$~T, ${\rm SNR}=1$, and including the effects of DM halo overdensities as well as velocities around the Earth and Jupiter (explained in Sec.~\ref{sec:jup-den}). 
Using Eq.~(\ref{eq:snr_1}), we derive the projected sensitivity reaches for the proposed experiment, onboard space missions on the surface of Jupiter (as a demonstration), at the location of IO, Callisto, as well as probed by the same experimental setup on Earth. These results are presented in Fig.~\ref{fig:dem}.  
In the figure, the dashed purple curve indicates the naturalness requirement on the couplings at a cutoff scale of $\Lambda = 3~\text{TeV}$, while the dotted purple curve marks the threshold at which the relaxion mechanism via the Higgs Portal remains physically valid~\cite{Flacke:2016szy,Choi:2016luu}.
We have indicated in our analysis that upcoming space-based quantum-sensor experiments have the potential to study the benchmark relaxion model for electron and photon couplings. 
However, laboratory experiments on Earth would require more advancements in their sensitivity to reach detection capabilities comparable to those in space.  These terrestrial experiments should be improved by several orders of magnitude,  particularly within specific regions of the mass parameter space.

\section{Conclusions}
\label{sec:conclusion}

We propose the use of space-based quantum sensors to study ultralight dark matter bound to planetary bodies, with a focus on Jupiter as a demonstration platform and optimal target due to its strong gravitational field. Leveraging data from missions such as JUNO and JUICE, we explored dark matter detection via gravitational and electromagnetic interactions. Jupiter’s mass makes it a prime candidate for high-concentration dark matter studies, providing a unique experimental setting.

Our analysis builds on the concept of utilizing a magnetometer experiment with a permanent magnet onboard a spacecraft to detect oscillating magnetic fields induced by dark matter interactions. By employing precision magnetometers in the spacecraft environment near Jupiter, we aimed to place stringent constraints on dark matter’s mass and couplings, particularly within the frequency range of $1 \, \mathrm{Hz}$ to $100 \, \mathrm{kHz}$, corresponding to the mass of ultralight dark matter candidates.

We also considered bounded ULDM around Jupiter, and proposed theoretical bounds on dark matter properties. Combining magnetometers with quantum sensors offers an opportunity to detect coherent dark matter halos gravitationally bound to Jupiter. This interdisciplinary approach suggests that upcoming space missions could significantly improve constraints on dark matter-photon and dark matter-electron interactions, enhancing the detection prospects for ultralight dark matter in the solar system. Future investigations will refine this setup for practical implementation in Jupiter’s exploration.

\section*{Acknowledgements}

We would also like to thank Joseph Levine, Kuver Sinha, and Tao Xu for their useful discussions during the early stages of this project.  
F.~H. is supported by Homer Dodge postdoctoral fellowship and in part by DOE grant DE-SC0009956. 
F.H. also thanks the organizers of the workshop of the Center for Theoretical Underground Physics and Related Areas (CETUP* - 2024), The Institute for Underground Science at Sanford Underground Research Facility (SURF), Lead, South Dakota, for their hospitality and financial support.  Moreover, F.~H. thanks the organizers of the Mitchell Conference in May 2024 at Texas A \& M University for their hospitality and support during the initial stages of this project. This research is partially supported by LANL's Laboratory Directed Research and Development (LDRD) program. Y. D.-T. thanks the generous support from the LANL Director's Fellowship.
This research is partially supported by the U.S. National Science Foundation (NSF) Theoretical Physics Program, Grant No.~PHY-1915005. This research was supported in part by grant NSF PHY-2309135 to the Kavli Institute for Theoretical Physics (KITP).
This work was partially performed at the Aspen Center for Physics, supported by National Science Foundation grant No.~PHY-2210452. This research was partly supported by the NSF under Grant No.~NSF PHY-1748958.  This document was partially prepared using the resources of the Fermi National Accelerator Laboratory (Fermilab), a U.S. Department of Energy, Office of Science, HEP User Facility. Fermilab is managed by Fermi Research Alliance, LLC (FRA), acting under Contract No. DE-AC02-07CH11359.

\appendix
\section{{Details of Jupiter Missions}}
\label{app:jupmis}

Several space missions, including JUNO, JUICE, Europa Clipper, and Io Volcano Observer, offer unique opportunities to explore dark matter (DM) in Jupiter’s environment. The mass of Jupiter ($\sim$318 Earth masses) and its strong gravitational field make it an ideal target for DM studies. In particular, it is useful for detecting potential DM overdensities. If advanced quantum sensors such as atomic clocks and magnetometers are included in these missions, they could probe gravitational and electromagnetic effects induced by ULDM interactions.

The JUNO mission, which has been operational since 2016, has provided valuable data on the structure of Jupiter, its magnetic field, and atmospheric dynamics \cite{bolton2017jupiter, matousek2007JUNO,connerney2017jupiter, adriani2018clusters}. It has also been used to set constraints on dark photon models \cite{Yan:2023kdg}. The JUICE mission, launched in 2023 and set to reach Jupiter in 2031, will conduct extensive studies of the habitability of Ganymede and perform multiple flybys of Europa and Callisto \cite{grasset2013jupiter, witasse2021juice, grasset2013review, dougherty2012jupiter}. Meanwhile, Europa Clipper, which has been launched in 2024, will investigate the ice shell of Europa and its subsurface ocean to look for its potential for life \cite{bayer2015europa,phillips2014europa, pappalardo2013science, pappalardo2024science, vance2023investigating}. A proposed Europa Lander (possibly launching in the 2030s) intends to analyze surface ice and subsurface water for biological signatures and chemical composition \cite{nasa2024search, wall2024maps, nasa2016europa, dooley2018mission,cameron2021europa,hoey2020europa,hand2022science}.

The Io Volcano Observer is planned to study Io volcano activities and its interactions with Jupiter’s magnetosphere \cite{hamilton2024comparing,bhardwaj2002i0,de2024isotopic,davies2024io,tyler2015tidal}. Different Jupiter missions, along with future proposals, could provide opportunities for novel DM detection techniques such as atomic clocks and quantum sensors to improve constraints on DM mass and interactions, which is the main focus of this paper. Future missions like Dragonfly around the moons of Saturn, may provide better constraints \cite{lorenz2018dragonfly, barnes2021science, lorenz2021engineering, trainer2018dragonfly,nasa_dragonfly}, which we leave for future studies. If DM-sensitive experiments are included in planned planetary missions, they can advance  DM searches within the solar system.

Tables \ref{table:jupiter_moons} and \ref{tab:jupiter_missions} provide further details on Jupiter’s moons and relevant missions, highlighting their potential applications for DM studies.

 \paragraph*{\rm Note Added:}
 The concept and preliminary results of this paper were presented on May 15, 2024~\cite{Tsai_NASA24}  and November 3, 2024~\cite{Hajkarim_A24} by the authors.
 While we were preparing this manuscript, another study appeared using a different experimental approach and planetary targets to search for fast-oscillating scalar fields~\cite{Budker:2024bzj}.

\begin{table*}[!]
\centering
\setlength{\extrarowheight}{1.5mm}
\begin{tabular}{|c|c|c|c|}
\hline
\textbf{Mission} & \textbf{Year of Start and End} & \textbf{Total Length of Mission} & \textbf{Time Spent Around Each Moon} \\
\hline
JUICE & 2023 -- 2035& $\sim$ 12 years & \begin{tabular}{@{}c@{}}Ganymede: $\sim$ 9 months \\ Europa: $\sim$ a few days \\ Callisto: $\sim$ 2.5 years
\\ Jupiter: $\sim$ 3.5 years \end{tabular} \\
\hline
JUNO & 2011 -- 2025 & $\sim$ 14 years & \begin{tabular}{@{}c@{}}Jupiter: $\sim$ 20 months \\ (Focus on Ganymede, Europa, and \\ Io during close flybys) \end{tabular} \\
\hline
IVO & 2025 -- 2033 & $\sim$ 8 years & \begin{tabular}{@{}c@{}}Io: $\sim$ 4 years \\ (Primarily focused on Io) \end{tabular} \\
\hline
Europa Clipper & 2024 -- 2034 & $\sim$ 10 years & \begin{tabular}{@{}c@{}}Europa: $\sim$ 4 years \\ (Multiple flybys of Europa) \end{tabular} \\
\hline
Europa Lander & 2025 -- 2032 & $\sim$ 7 years & \begin{tabular}{@{}c@{}}Europa: $\sim$ 1 month (on surface) \\ (Primarily focused on surface operations) \end{tabular} \\
\hline
\end{tabular}

\caption{Overview of missions near Jupiter and its moons with their respective start and end years \cite{grasset2013jupiter, witasse2021juice, grasset2013review, dougherty2012jupiter, bolton2017jupiter, matousek2007JUNO, connerney2017jupiter, adriani2018clusters, mcewen2014io, williams2011volcanism, phillips2014europa, pappalardo2013science, pappalardo2024science, vance2023investigating, pappalardo2024science}.}
\label{tab:jupiter_missions}
\end{table*}

\bibliographystyle{JHEP}
\bibliography{ref}

@inproceedings{Baryakhtar:2022hbu,
    author = "Baryakhtar, Masha and others",
    title = "{Dark Matter In Extreme Astrophysical Environments}",
    booktitle = "{Snowmass 2021}",
    eprint = "2203.07984",
    archivePrefix = "arXiv",
    primaryClass = "hep-ph",
    month = "3",
    year = "2022"
}

@inproceedings{Adams:2022pbo,
    author = "Adams, C. B. and others",
    title = "{Axion Dark Matter}",
    booktitle = "{Snowmass 2021}",
    eprint = "2203.14923",
    archivePrefix = "arXiv",
    primaryClass = "hep-ex",
    reportNumber = "FERMILAB-CONF-22-996-PPD-T",
    month = "3",
    year = "2022"
}

@article{Green:2022hhj,
    author = "Green, Daniel and others",
    title = "{Snowmass Theory Frontier: Astrophysics and Cosmology}",
    eprint = "2209.06854",
    archivePrefix = "arXiv",
    primaryClass = "hep-ph",
    reportNumber = "FERMILAB-PUB-22-721-T",
    month = "9",
    year = "2022"
}

@article{Feng:2010gw,
    author = "Feng, Jonathan L.",
    title = "{Dark Matter Candidates from Particle Physics and Methods of Detection}",
    eprint = "1003.0904",
    archivePrefix = "arXiv",
    primaryClass = "astro-ph.CO",
    reportNumber = "UCI-TR-2009-13",
    doi = "10.1146/annurev-astro-082708-101659",
    journal = "Ann. Rev. Astron. Astrophys.",
    volume = "48",
    pages = "495--545",
    year = "2010"
}

@article{Budker:2013hfa,
    author = "Budker, Dmitry and Graham, Peter W. and Ledbetter, Micah and Rajendran, Surjeet and Sushkov, Alex",
    title = "{Proposal for a Cosmic Axion Spin Precession Experiment (CASPEr)}",
    eprint = "1306.6089",
    archivePrefix = "arXiv",
    primaryClass = "hep-ph",
    doi = "10.1103/PhysRevX.4.021030",
    journal = "Phys. Rev. X",
    volume = "4",
    number = "2",
    pages = "021030",
    year = "2014"
}

@article{Graham:2015ifn,
    author = "Graham, Peter W. and Kaplan, David E. and Mardon, Jeremy and Rajendran, Surjeet and Terrano, William A.",
    title = "{Dark Matter Direct Detection with Accelerometers}",
    eprint = "1512.06165",
    archivePrefix = "arXiv",
    primaryClass = "hep-ph",
    doi = "10.1103/PhysRevD.93.075029",
    journal = "Phys. Rev. D",
    volume = "93",
    number = "7",
    pages = "075029",
    year = "2016"
}

@article{Antypas:2019qji,
    author = "Antypas, D. and Tretiak, O. and Garcon, A. and Ozeri, R. and Perez, G. and Budker, D.",
    title = "{Scalar dark matter in the radio-frequency band: atomic-spectroscopy search results}",
    eprint = "1905.02968",
    archivePrefix = "arXiv",
    primaryClass = "physics.atom-ph",
    doi = "10.1103/PhysRevLett.123.141102",
    journal = "Phys. Rev. Lett.",
    volume = "123",
    number = "14",
    pages = "141102",
    year = "2019"
}

@article{Gramolin:2020ict,
    author = "Gramolin, Alexander V. and Aybas, Deniz and Johnson, Dorian and Adam, Janos and Sushkov, Alexander O.",
    title = "{Search for axion-like dark matter with ferromagnets}",
    eprint = "2003.03348",
    archivePrefix = "arXiv",
    primaryClass = "hep-ex",
    doi = "10.1038/s41567-020-1006-6",
    journal = "Nature Phys.",
    volume = "17",
    number = "1",
    pages = "79--84",
    year = "2021"
}

@article{Oswald:2021vtc,
    author = "Oswald, R. and others",
    title = "{Search for Dark-Matter-Induced Oscillations of Fundamental Constants Using Molecular Spectroscopy}",
    eprint = "2111.06883",
    archivePrefix = "arXiv",
    primaryClass = "hep-ph",
    doi = "10.1103/PhysRevLett.129.031302",
    journal = "Phys. Rev. Lett.",
    volume = "129",
    number = "3",
    pages = "031302",
    year = "2022"
}

@article{Antypas:2022asj,
    author = "Antypas, D. and others",
    title = "{New Horizons: Scalar and Vector Ultralight Dark Matter}",
    eprint = "2203.14915",
    archivePrefix = "arXiv",
    primaryClass = "hep-ex",
    reportNumber = "FERMILAB-PUB-22-262-AD-PPD-T",
    month = "3",
    year = "2022"
}

@article{Tretiak:2022ndx,
    author = "Tretiak, Oleg and Zhang, Xue and Figueroa, Nataniel L. and Antypas, Dionysios and Brogna, Andrea and Banerjee, Abhishek and Perez, Gilad and Budker, Dmitry",
    title = "{Improved Bounds on Ultralight Scalar Dark Matter in the Radio-Frequency Range}",
    eprint = "2201.02042",
    archivePrefix = "arXiv",
    primaryClass = "hep-ph",
    doi = "10.1103/PhysRevLett.129.031301",
    journal = "Phys. Rev. Lett.",
    volume = "129",
    number = "3",
    pages = "031301",
    year = "2022"
}

@article{Bloch:2023wfz,
    author = "Bloch, Itay M. and Kalia, Saarik",
    title = "{Curl up with a good B: detecting ultralight dark matter with differential magnetometry}",
    eprint = "2308.10931",
    archivePrefix = "arXiv",
    primaryClass = "hep-ph",
    reportNumber = "FERMILAB-PUB-23-498-SQMS-V",
    doi = "10.1007/JHEP01(2024)178",
    journal = "JHEP",
    volume = "2024",
    number = "1",
    pages = "178",
    year = "2024"
}

@article{Bloch:2023uis,
    author = "Bloch, I. M. and Budker, D. and Flambaum, V. V. and Samsonov, I. B. and Sushkov, A. O. and Tretiak, O.",
    title = "{Scalar dark matter induced oscillation of a permanent-magnet field}",
    eprint = "2301.08514",
    archivePrefix = "arXiv",
    primaryClass = "hep-ph",
    doi = "10.1103/PhysRevD.107.075033",
    journal = "Phys. Rev. D",
    volume = "107",
    number = "7",
    pages = "075033",
    year = "2023"
}

@article{Poddar:2019zoe,
    author = "Kumar Poddar, Tanmay and Mohanty, Subhendra and Jana, Soumya",
    title = "{Constraints on ultralight axions from compact binary systems}",
    eprint = "1906.00666",
    archivePrefix = "arXiv",
    primaryClass = "hep-ph",
    doi = "10.1103/PhysRevD.101.083007",
    journal = "Phys. Rev. D",
    volume = "101",
    number = "8",
    pages = "083007",
    year = "2020"
}

@article{Tsai:2023zza,
    author = "Tsai, Yu-Dai and Farnocchia, Davide and Micheli, Marco and Vagnozzi, Sunny and Visinelli, Luca",
    title = "{Constraints on fifth forces and ultralight dark matter from OSIRIS-REx target asteroid Bennu}",
    eprint = "2309.13106",
    archivePrefix = "arXiv",
    primaryClass = "hep-ph",
    reportNumber = "UCI-HEP-TR-2023-04, FERMILAB-PUB-23-538-T-V",
    doi = "10.1038/s42005-024-01779-3",
    journal = "Commun. Phys.",
    volume = "7",
    number = "1",
    pages = "311",
    year = "2024"
}

@article{Bar:2019bqz,
    author = "Bar, Nitsan and Blum, Kfir and Eby, Joshua and Sato, Ryosuke",
    title = "{Ultralight dark matter in disk galaxies}",
    eprint = "1903.03402",
    archivePrefix = "arXiv",
    primaryClass = "astro-ph.CO",
    reportNumber = "DESY-19-036",
    doi = "10.1103/PhysRevD.99.103020",
    journal = "Phys. Rev. D",
    volume = "99",
    number = "10",
    pages = "103020",
    year = "2019"
}

@article{Tsai:2021lly,
    author = "Tsai, Yu-Dai and Eby, Joshua and Safronova, Marianna S.",
    title = "{Direct detection of ultralight dark matter bound to the Sun with space quantum sensors}",
    eprint = "2112.07674",
    archivePrefix = "arXiv",
    primaryClass = "hep-ph",
    reportNumber = "FERMILAB-PUB-21-687-T-V, IPMU21-0085",
    doi = "10.1038/s41550-022-01833-6",
    journal = "Nature Astron.",
    volume = "7",
    number = "1",
    pages = "113--121",
    year = "2023"
}

@article{Tsai:2020vpi,
    author = "Tsai, Yu-Dai and McGehee, Robert and Murayama, Hitoshi",
    title = "{Resonant Self-Interacting Dark Matter from Dark QCD}",
    eprint = "2008.08608",
    archivePrefix = "arXiv",
    primaryClass = "hep-ph",
    reportNumber = "FERMILAB-PUB-20-365-AE-T",
    doi = "10.1103/PhysRevLett.128.172001",
    journal = "Phys. Rev. Lett.",
    volume = "128",
    number = "17",
    pages = "172001",
    year = "2022"
}

@inbook{JacksonKimball:2023ers,
    author = "Jackson Kimball, Derek F. and Phipps, Arran",
    editor = "Kimball, Derek F. Jackson and van Bibber, Karl",
    title = "{Dark Matter Radios}",
    booktitle = "{The Search for Ultralight Bosonic Dark Matter}",
    doi = "10.1007/978-3-030-95852-7_7",
    pages = "201--218",
    year = "2023"
}

@article{Kennedy:2020bac,
    author = "Kennedy, Colin J. and Oelker, Eric and Robinson, John M. and Bothwell, Tobias and Kedar, Dhruv and Milner, William R. and Marti, G. Edward and Derevianko, Andrei and Ye, Jun",
    title = "{Precision Metrology Meets Cosmology: Improved Constraints on Ultralight Dark Matter from Atom-Cavity Frequency Comparisons}",
    eprint = "2008.08773",
    archivePrefix = "arXiv",
    primaryClass = "physics.atom-ph",
    doi = "10.1103/PhysRevLett.125.201302",
    journal = "Phys. Rev. Lett.",
    volume = "125",
    number = "20",
    pages = "201302",
    year = "2020"
}

@article{Sibiryakov:2020eir,
    author = "Sibiryakov, Sergey and Sorensen, Philip and Yu, Tien-Tien",
    title = "{BBN constraints on universally-coupled ultralight scalar dark matter}",
    eprint = "2006.04820",
    archivePrefix = "arXiv",
    primaryClass = "hep-ph",
    reportNumber = "DESY-19-234, CERN-TH-2020-091, INR-TH-2020-001",
    doi = "10.1007/JHEP12(2020)075",
    journal = "JHEP",
    volume = "12",
    pages = "075",
    year = "2020"
}

@article{Hui:2016ltb,
    author = "Hui, Lam and Ostriker, Jeremiah P. and Tremaine, Scott and Witten, Edward",
    title = "{Ultralight scalars as cosmological dark matter}",
    eprint = "1610.08297",
    archivePrefix = "arXiv",
    primaryClass = "astro-ph.CO",
    doi = "10.1103/PhysRevD.95.043541",
    journal = "Phys. Rev. D",
    volume = "95",
    number = "4",
    pages = "043541",
    year = "2017"
}

@article{Baryakhtar:2017ngi,
    author = "Baryakhtar, Masha and Lasenby, Robert and Teo, Mae",
    title = "{Black Hole Superradiance Signatures of Ultralight Vectors}",
    eprint = "1704.05081",
    archivePrefix = "arXiv",
    primaryClass = "hep-ph",
    doi = "10.1103/PhysRevD.96.035019",
    journal = "Phys. Rev. D",
    volume = "96",
    number = "3",
    pages = "035019",
    year = "2017"
}

@article{Hlozek:2014lca,
    author = "Hlozek, Ren\'ee and Grin, Daniel and Marsh, David J. E. and Ferreira, Pedro G.",
    title = "{A search for ultralight axions using precision cosmological data}",
    eprint = "1410.2896",
    archivePrefix = "arXiv",
    primaryClass = "astro-ph.CO",
    doi = "10.1103/PhysRevD.91.103512",
    journal = "Phys. Rev. D",
    volume = "91",
    number = "10",
    pages = "103512",
    year = "2015"
}

@article{Davoudiasl:2019nlo,
    author = "Davoudiasl, Hooman and Denton, Peter B",
    title = "{Ultralight Boson Dark Matter and Event Horizon Telescope Observations of M87*}",
    eprint = "1904.09242",
    archivePrefix = "arXiv",
    primaryClass = "astro-ph.CO",
    doi = "10.1103/PhysRevLett.123.021102",
    journal = "Phys. Rev. Lett.",
    volume = "123",
    number = "2",
    pages = "021102",
    year = "2019"
}

@article{Roy:2019esk,
    author = "Roy, Rittick and Yajnik, Urjit A.",
    title = "{Evolution of black hole shadow in the presence of ultralight bosons}",
    eprint = "1906.03190",
    archivePrefix = "arXiv",
    primaryClass = "gr-qc",
    doi = "10.1016/j.physletb.2020.135284",
    journal = "Phys. Lett. B",
    volume = "803",
    pages = "135284",
    year = "2020"
}

@article{Vagnozzi:2022moj,
    author = "Vagnozzi, Sunny and others",
    title = "{Horizon-scale tests of gravity theories and fundamental physics from the Event Horizon Telescope image of Sagittarius A}",
    eprint = "2205.07787",
    archivePrefix = "arXiv",
    primaryClass = "gr-qc",
    reportNumber = "UCI-HEP-TR-2022-07",
    doi = "10.1088/1361-6382/acd97b",
    journal = "Class. Quant. Grav.",
    volume = "40",
    number = "16",
    pages = "165007",
    year = "2023"
}

@article{Gau:2023rct,
    author = "Gau, Ephraim and Hajkarim, Fazlollah and Harris, Steven P. and Dev, P. S. Bhupal and Fortin, Jean-Francois and Krawczynski, Henric and Sinha, Kuver",
    title = "{New constraints on axion-like particles from IXPE polarization data for magnetars}",
    eprint = "2312.14153",
    archivePrefix = "arXiv",
    primaryClass = "hep-ph",
    reportNumber = "INT-PUB-23-053",
    doi = "10.1016/j.dark.2024.101709",
    journal = "Phys. Dark Univ.",
    volume = "46",
    pages = "101709",
    year = "2024"
}

@article{Shakeri:2022usk,
    author = "Shakeri, Soroush and Hajkarim, Fazlollah",
    title = "{Probing axions via light circular polarization and event horizon telescope}",
    eprint = "2209.13572",
    archivePrefix = "arXiv",
    primaryClass = "hep-ph",
    doi = "10.1088/1475-7516/2023/04/017",
    journal = "JCAP",
    volume = "04",
    pages = "017",
    year = "2023"
}

@article{Fortin:2018aom,
    author = "Fortin, Jean-Fran\c{c}ois and Sinha, Kuver",
    title = "{X-Ray Polarization Signals from Magnetars with Axion-Like-Particles}",
    eprint = "1807.10773",
    archivePrefix = "arXiv",
    primaryClass = "hep-ph",
    doi = "10.1007/JHEP01(2019)163",
    journal = "JHEP",
    volume = "01",
    pages = "163",
    year = "2019"
}

@article{Yan:2023kdg,
    author = "Yan, Shi and Li, Lingfeng and Fan, JiJi",
    title = "{Constraints on photon mass and dark photon from the Jovian magnetic field}",
    eprint = "2312.06746",
    archivePrefix = "arXiv",
    primaryClass = "hep-ph",
    doi = "10.1007/JHEP06(2024)028",
    journal = "JHEP",
    volume = "06",
    pages = "028",
    year = "2024"
}

@article{Banerjee:2019epw,
    author = "Banerjee, Abhishek and Budker, Dmitry and Eby, Joshua and Kim, Hyungjin and Perez, Gilad",
    title = "{Relaxion Stars and their detection via Atomic Physics}",
    eprint = "1902.08212",
    archivePrefix = "arXiv",
    primaryClass = "hep-ph",
    doi = "10.1038/s42005-019-0260-3",
    journal = "Commun. Phys.",
    volume = "3",
    pages = "1",
    year = "2020"
}

@article{Hees:2018fpg,
    author = "Hees, Aur\'elien and Minazzoli, Olivier and Savalle, Etienne and Stadnik, Yevgeny V. and Wolf, Peter",
    title = "{Violation of the equivalence principle from light scalar dark matter}",
    eprint = "1807.04512",
    archivePrefix = "arXiv",
    primaryClass = "gr-qc",
    doi = "10.1103/PhysRevD.98.064051",
    journal = "Phys. Rev. D",
    volume = "98",
    number = "6",
    pages = "064051",
    year = "2018"
}

@article{Tsai:2022jnv,
    author = "Tsai, Yu-Dai and Eby, Joshua and Arakawa, Jason and Farnocchia, Davide and Safronova, Marianna S.",
    title = "{OSIRIS-REx constraints on local dark matter and cosmic neutrino profiles}",
    eprint = "2210.03749",
    archivePrefix = "arXiv",
    primaryClass = "hep-ph",
    reportNumber = "UCI-HEP-TR-2022-11, FERMILAB-PUB-22-753-T-V",
    doi = "10.1088/1475-7516/2024/02/029",
    journal = "JCAP",
    volume = "02",
    pages = "029",
    year = "2024"
}

@article{Tsai:2021irw,
    author = "Tsai, Yu-Dai and Wu, Youjia and Vagnozzi, Sunny and Visinelli, Luca",
    title = "{Novel constraints on fifth forces and ultralight dark sector with asteroidal data}",
    eprint = "2107.04038",
    archivePrefix = "arXiv",
    primaryClass = "hep-ph",
    reportNumber = "FERMILAB-PUB-21-298-AE-T, LCTP-21-17",
    doi = "10.1088/1475-7516/2023/04/031",
    journal = "JCAP",
    volume = "04",
    pages = "031",
    year = "2023"
}

@article{Bertone:2004pz,
    author = "Bertone, Gianfranco and Hooper, Dan and Silk, Joseph",
    title = "{Particle dark matter: Evidence, candidates and constraints}",
    eprint = "hep-ph/0404175",
    archivePrefix = "arXiv",
    reportNumber = "FERMILAB-PUB-04-047-A",
    doi = "10.1016/j.physrep.2004.08.031",
    journal = "Phys. Rept.",
    volume = "405",
    pages = "279--390",
    year = "2005"
}

@article{Cirelli:2024ssz,
    author = "Cirelli, Marco and Strumia, Alessandro and Zupan, Jure",
    title = "{Dark Matter}",
    eprint = "2406.01705",
    archivePrefix = "arXiv",
    primaryClass = "hep-ph",
    month = "6",
    year = "2024"
}

@article{Banerjee:2019xuy,
    author = "Banerjee, Abhishek and Budker, Dmitry and Eby, Joshua and Flambaum, Victor V. and Kim, Hyungjin and Matsedonskyi, Oleksii and Perez, Gilad",
    title = "{Searching for Earth/Solar Axion Halos}",
    eprint = "1912.04295",
    archivePrefix = "arXiv",
    primaryClass = "hep-ph",
    doi = "10.1007/JHEP09(2020)004",
    journal = "JHEP",
    volume = "09",
    pages = "004",
    year = "2020"
}

@article{Budker:2023sex,
    author = "Budker, Dmitry and Eby, Joshua and Gorghetto, Marco and Jiang, Minyuan and Perez, Gilad",
    title = "{A generic formation mechanism of ultralight dark matter solar halos}",
    eprint = "2306.12477",
    archivePrefix = "arXiv",
    primaryClass = "hep-ph",
    doi = "10.1088/1475-7516/2023/12/021",
    journal = "JCAP",
    volume = "12",
    pages = "021",
    year = "2023"
}

@article{Pitjev:2013sfa,
    author = "Pitjev, N. P. and Pitjeva, E. V.",
    title = "{Constraints on dark matter in the solar system}",
    eprint = "1306.5534",
    archivePrefix = "arXiv",
    primaryClass = "astro-ph.EP",
    doi = "10.1134/S1063773713020060",
    journal = "Astron. Lett.",
    volume = "39",
    pages = "141--149",
    year = "2013"
}

@article{Adler:2008rq,
    author = "Adler, Stephen L.",
    title = "{Placing direct limits on the mass of earth-bound dark matter}",
    eprint = "0808.0899",
    archivePrefix = "arXiv",
    primaryClass = "astro-ph",
    doi = "10.1088/1751-8113/41/41/412002",
    journal = "J. Phys. A",
    volume = "41",
    pages = "412002",
    year = "2008"
}

@article{DiLuzio:2020wdo,
    author = "Di Luzio, Luca and Giannotti, Maurizio and Nardi, Enrico and Visinelli, Luca",
    title = "{The landscape of QCD axion models}",
    eprint = "2003.01100",
    archivePrefix = "arXiv",
    primaryClass = "hep-ph",
    reportNumber = "DESY 20-036, DESY-20-036",
    doi = "10.1016/j.physrep.2020.06.002",
    journal = "Phys. Rept.",
    volume = "870",
    pages = "1--117",
    year = "2020"
}

@article{Sherrill:2023zah,
    author = "Sherrill, Nathaniel and others",
    title = "{Analysis of atomic-clock data to constrain variations of fundamental constants}",
    eprint = "2302.04565",
    archivePrefix = "arXiv",
    primaryClass = "physics.atom-ph",
    doi = "10.1088/1367-2630/aceff6",
    journal = "New J. Phys.",
    volume = "25",
    number = "9",
    pages = "093012",
    year = "2023"
}

@article{Arvanitaki:2015iga,
    author = "Arvanitaki, Asimina and Dimopoulos, Savas and Van Tilburg, Ken",
    title = "{Sound of Dark Matter: Searching for Light Scalars with Resonant-Mass Detectors}",
    eprint = "1508.01798",
    archivePrefix = "arXiv",
    primaryClass = "hep-ph",
    doi = "10.1103/PhysRevLett.116.031102",
    journal = "Phys. Rev. Lett.",
    volume = "116",
    number = "3",
    pages = "031102",
    year = "2016"
}

@article{Fosbinder-Elkins:2017osp,
    author = "Fosbinder-Elkins, H. and Dargert, J. and Harkness, M. and Geraci, A. A. and Levenson-Falk, E. and Mumford, S. and Kapitulnik, A. and Shin, Y. and Semertzidis, Y. and Lee, Y. -H.",
    title = "{A method for controlling the magnetic field near a superconducting boundary in the ARIADNE axion experiment}",
    eprint = "1710.08102",
    archivePrefix = "arXiv",
    primaryClass = "physics.ins-det",
    month = "10",
    year = "2017"
}

@article{Li:2022wix,
    author = "Li, Lingfeng and Fan, JiJi",
    title = "{Jupiter missions as probes of dark matter}",
    eprint = "2207.13709",
    archivePrefix = "arXiv",
    primaryClass = "hep-ph",
    doi = "10.1007/JHEP10(2022)186",
    journal = "JHEP",
    volume = "10",
    pages = "186",
    year = "2022"
}

@article{Morgante:2009dqw,
    author = "Morgante, G. and others",
    title = "{Cryogenic characterization of the Planck sorption cooler system flight model}",
    eprint = "1001.4628",
    archivePrefix = "arXiv",
    primaryClass = "astro-ph.IM",
    doi = "10.1088/1748-0221/4/12/T12016",
    journal = "JINST",
    volume = "4",
    pages = "T12016",
    year = "2009"
}

@article{LiteBIRD:2022cnt,
    author = "Allys, E. and others",
    collaboration = "LiteBIRD",
    title = "{Probing Cosmic Inflation with the LiteBIRD Cosmic Microwave Background Polarization Survey}",
    eprint = "2202.02773",
    archivePrefix = "arXiv",
    primaryClass = "astro-ph.IM",
    doi = "10.1093/ptep/ptac150",
    journal = "PTEP",
    volume = "2023",
    number = "4",
    pages = "042F01",
    year = "2023"
}

@article{Matsumura:2013aja,
    author = "Matsumura, T. and others",
    title = "{Mission design of LiteBIRD}",
    eprint = "1311.2847",
    archivePrefix = "arXiv",
    primaryClass = "astro-ph.IM",
    doi = "10.1007/s10909-013-0996-1",
    journal = "J. Low Temp. Phys.",
    volume = "176",
    pages = "733",
    year = "2014"
}

@article{Suzuki:2018cuy,
    author = "Suzuki, A. and others",
    title = "{The LiteBIRD Satellite Mission - Sub-Kelvin Instrument}",
    eprint = "1801.06987",
    archivePrefix = "arXiv",
    primaryClass = "astro-ph.IM",
    doi = "10.1007/s10909-018-1947-7",
    journal = "J. Low Temp. Phys.",
    volume = "193",
    number = "5-6",
    pages = "1048--1056",
    year = "2018"
}

@article{LiteBIRD:2020zfx,
    author = "Montier, L. and others",
    collaboration = "LiteBIRD",
    title = "{Overview of the Medium and High Frequency Telescopes of the LiteBIRD satellite mission}",
    eprint = "2102.00809",
    archivePrefix = "arXiv",
    primaryClass = "astro-ph.IM",
    doi = "10.1117/12.2562243",
    journal = "Proc. SPIE Int. Soc. Opt. Eng.",
    volume = "11443",
    pages = "114432G",
    year = "2020"
}

@article{Gardner:2006ky,
    author = "Gardner, Jonathan P. and others",
    title = "{The James Webb Space Telescope}",
    eprint = "astro-ph/0606175",
    archivePrefix = "arXiv",
    doi = "10.1007/s11214-006-8315-7",
    journal = "Space Sci. Rev.",
    volume = "123",
    pages = "485",
    year = "2006"
}

@article{Chaudhuri:2014dla,
    author = "Chaudhuri, Saptarshi and Graham, Peter W. and Irwin, Kent and Mardon, Jeremy and Rajendran, Surjeet and Zhao, Yue",
    title = "{Radio for hidden-photon dark matter detection}",
    eprint = "1411.7382",
    archivePrefix = "arXiv",
    primaryClass = "hep-ph",
    doi = "10.1103/PhysRevD.92.075012",
    journal = "Phys. Rev. D",
    volume = "92",
    number = "7",
    pages = "075012",
    year = "2015"
}

@article{JacksonKimball:2016wzv,
    author = "Jackson Kimball, D. F. and Dudley, J. and Li, Y. and Thulasi, S. and Pustelny, S. and Budker, D. and Zolotorev, M.",
    title = "{Magnetic shielding and exotic spin-dependent interactions}",
    eprint = "1606.00696",
    archivePrefix = "arXiv",
    primaryClass = "physics.ins-det",
    doi = "10.1103/PhysRevD.94.082005",
    journal = "Phys. Rev. D",
    volume = "94",
    number = "8",
    pages = "082005",
    year = "2016"
}

@article{Flacke:2016szy,
    author = "Flacke, Thomas and Frugiuele, Claudia and Fuchs, Elina and Gupta, Rick S. and Perez, Gilad",
    title = "{Phenomenology of relaxion-Higgs mixing}",
    eprint = "1610.02025",
    archivePrefix = "arXiv",
    primaryClass = "hep-ph",
    reportNumber = "CTPU-16-25",
    doi = "10.1007/JHEP06(2017)050",
    journal = "JHEP",
    volume = "06",
    pages = "050",
    year = "2017"
}

@article{Choi:2016luu,
    author = "Choi, Kiwoon and Im, Sang Hui",
    title = "{Constraints on Relaxion Windows}",
    eprint = "1610.00680",
    archivePrefix = "arXiv",
    primaryClass = "hep-ph",
    reportNumber = "CTPU-16-29",
    doi = "10.1007/JHEP12(2016)093",
    journal = "JHEP",
    volume = "12",
    pages = "093",
    year = "2016"
}

@article{Wagner:2012ui,
    author = "Wagner, T. A. and Schlamminger, S. and Gundlach, J. H. and Adelberger, E. G.",
    title = "{Torsion-balance tests of the weak equivalence principle}",
    eprint = "1207.2442",
    archivePrefix = "arXiv",
    primaryClass = "gr-qc",
    doi = "10.1088/0264-9381/29/18/184002",
    journal = "Class. Quant. Grav.",
    volume = "29",
    pages = "184002",
    year = "2012"
}

@article{Berge:2017ovy,
    author = "Berg\'e, Joel and Brax, Philippe and M\'etris, Gilles and Pernot-Borr\`as, Martin and Touboul, Pierre and Uzan, Jean-Philippe",
    title = "{MICROSCOPE Mission: First Constraints on the Violation of the Weak Equivalence Principle by a Light Scalar Dilaton}",
    eprint = "1712.00483",
    archivePrefix = "arXiv",
    primaryClass = "gr-qc",
    doi = "10.1103/PhysRevLett.120.141101",
    journal = "Phys. Rev. Lett.",
    volume = "120",
    number = "14",
    pages = "141101",
    year = "2018"
}

@article{Antypas:2020rtg,
    author = "Antypas, Dionysios and Tretiak, Oleg and Zhang, Ke and Garcon, Antoine and Perez, Gilad and Kozlov, Mikhail G. and Schiller, Stephan and Budker, Dmitry",
    title = "{Probing fast oscillating scalar dark matter with atoms and molecules}",
    eprint = "2012.01519",
    archivePrefix = "arXiv",
    primaryClass = "physics.atom-ph",
    doi = "10.1088/2058-9565/abe472",
    journal = "Quantum Sci. Technol.",
    volume = "6",
    number = "3",
    pages = "034001",
    year = "2021"
}

@article{pitjev2013constraints,
  title={Constraints on dark matter in the solar system},
  author={Pitjev, NP and Pitjeva, EV},
  journal={Astronomy Letters},
  volume={39},
  pages={141--149},
  year={2013},
  publisher={Springer}
}

@article{Blanco:2023qgi,
    author = "Blanco, Carlos and Leane, Rebecca K.",
    title = "{Search for Dark Matter Ionization on the Night Side of Jupiter with Cassini}",
    eprint = "2312.06758",
    archivePrefix = "arXiv",
    primaryClass = "hep-ph",
    reportNumber = "SLAC-PUB-17753",
    doi = "10.1103/PhysRevLett.132.261002",
    journal = "Phys. Rev. Lett.",
    volume = "132",
    number = "26",
    pages = "261002",
    year = "2024"
}

@book{bagenal2006jupiter,
  title={Jupiter: the planet, satellites and magnetosphere},
  author={Bagenal, Fran and Dowling, Timothy E and McKinnon, William B and McKinnon, William},
  volume={1},
  year={2006},
  publisher={Cambridge University Press}
}

@article{bolton2017jupiter,
  title={Jupiter’s interior and deep atmosphere: The initial pole-to-pole passes with the Juno spacecraft},
  author={Bolton, Scott J and Adriani, Alberto and Adumitroaie, V and Allison, M and Anderson, J and Atreya, S and Bloxham, J and Brown, S and Connerney, JEP and DeJong, E and others},
  journal={Science},
  volume={356},
  number={6340},
  pages={821--825},
  year={2017},
  publisher={American Association for the Advancement of Science}
}

@article{connerney2018new,
  title={A new model of Jupiter's magnetic field from Juno's first nine orbits},
  author={Connerney, JEP and Kotsiaros, S and Oliversen, RJ and Espley, JR and Joergensen, John Leif and Joergensen, PS and Merayo, Jos{\'e} MG and Herceg, Matija and Bloxham, J and Moore, KM and others},
  journal={Geophysical Research Letters},
  volume={45},
  number={6},
  pages={2590--2596},
  year={2018},
  publisher={Wiley Online Library}
}

@book{bagenal2007jupiter,
  title={Jupiter: The Planet, Satellites and Magnetosphere},
  author={Bagenal, Fran and Dowling, Timothy E. and McKinnon, William B.},
  year={2007},
  publisher={Cambridge University Press}
}

@book{cardwell2003handbook,
  title={Handbook of superconducting materials},
  author={Cardwell, David A and Ginley, David S},
  volume={1},
  year={2003},
  publisher={Crc Press}
}

@book{mele2019superconductivity,
  title={Superconductivity: from materials science to practical applications},
  author={Mele, Paolo and Prassides, Kosmas and Tarantini, Chiara and Palau, Anna and Badica, Petre and Jha, Alok K and Endo, Tamio},
  year={2019},
  publisher={Springer Nature}
}

@book{buckel2008superconductivity,
  title={Superconductivity: fundamentals and applications},
  author={Buckel, Werner and Kleiner, Reinhold},
  year={2008},
  publisher={John Wiley \& Sons}
}

@article{aghanim2020planck,
  title={Planck 2018 results-I. Overview and the cosmological legacy of Planck},
  author={Aghanim, Nabila and Akrami, Yashar and Arroja, Frederico and Ashdown, Mark and Aumont, J and Baccigalupi, Carlo and Ballardini, M and Banday, Anthony J and Barreiro, RB and Bartolo, Nicola and others},
  journal={Astronomy \& Astrophysics},
  volume={641},
  pages={A1},
  year={2020},
  publisher={EDP sciences}
}

@article{lamarre2003planck,
  title={The Planck High Frequency Instrument, a third generation CMB experiment, and a full sky submillimeter survey},
  author={Lamarre, Jean Michael and Puget, JL and Bouchet, Freddy and Ade, Peter AR and Benoit, Ang{\'e}lique and Bernard, Jean-Paul and Bock, J and De Bernardis, Paolo and Charra, J and Couchot, F and others},
  journal={New Astronomy Reviews},
  volume={47},
  number={11-12},
  pages={1017--1024},
  year={2003},
  publisher={Elsevier}
}

@article{ressler2015mid,
  title={The mid-infrared instrument for the James Webb Space Telescope, VIII: the MIRI focal plane system},
  author={Ressler, ME and Sukhatme, KG and Franklin, BR and Mahoney, JC and Thelen, MP and Bouchet, P and Colbert, JW and Cracraft, Misty and Dicken, D and Gastaud, R and others},
  journal={Publications of the Astronomical Society of the Pacific},
  volume={127},
  number={953},
  pages={675},
  year={2015},
  publisher={IOP Publishing}
}

@article{lightsey2012james,
  title={James Webb Space Telescope: large deployable cryogenic telescope in space},
  author={Lightsey, Paul A and Atkinson, Charles and Clampin, Mark and Feinberg, Lee D},
  journal={Optical Engineering},
  volume={51},
  number={1},
  pages={011003--011003},
  year={2012},
  publisher={Society of Photo-Optical Instrumentation Engineers}
}

@article{gao2024analysis,
  title={Analysis and measurement of magnetic noise in multilayer magnetic shielding system combined with mu-metal and ferrite for atomic sensors},
  author={Gao, Yanan and Fang, Xiujie and Ma, Danyue and Sun, Bowen and Wang, Kun and Li, Siran and Dou, Yao and Zeng, Min},
  journal={Measurement},
  volume={233},
  pages={114745},
  year={2024},
  publisher={Elsevier}
}

@article{volegov2004simultaneous,
  title={Simultaneous magnetoencephalography and SQUID detected nuclear MR in microtesla magnetic fields},
  author={Volegov, Petr and Matlachov, Andrei N and Espy, Michelle A and George, John S and Kraus Jr, Robert H},
  journal={Magnetic Resonance in Medicine: An Official Journal of the International Society for Magnetic Resonance in Medicine},
  volume={52},
  number={3},
  pages={467--470},
  year={2004},
  publisher={Wiley Online Library}
}

@article{ma2020magnetic,
  title={Magnetic noise calculation of mu-metal shields at extremely low frequencies for atomic devices},
  author={Ma, Danyue and Ding, Ming and Lu, Jixi and Zhao, Junpeng and Yang, Ke and Fang, Xiujie and Wang, Kun and Zhang, Ning and Han, Bangcheng},
  journal={Journal of Physics D: Applied Physics},
  volume={54},
  number={2},
  pages={025004},
  year={2020},
  publisher={IOP Publishing}
}

@article{romalis2011atomic,
  title={Atomic magnetometers for materials characterization},
  author={Romalis, Michael V and Dang, Hoan B},
  journal={Materials today},
  volume={14},
  number={6},
  pages={258--262},
  year={2011},
  publisher={Elsevier}
}

@article{fang2022high,
  title={A high-performance magnetic shield with MnZn ferrite and Mu-metal film combination for atomic sensors},
  author={Fang, Xiujie and Ma, Danyue and Sun, Bowen and Xu, Xueping and Quan, Wei and Xiao, Zhisong and Zhai, Yueyang},
  journal={Materials},
  volume={15},
  number={19},
  pages={6680},
  year={2022},
  publisher={MDPI}
}

@article{matousek2007juno,
  title={The Juno new frontiers mission},
  author={Matousek, Steve},
  journal={Acta Astronautica},
  volume={61},
  number={10},
  pages={932--939},
  year={2007},
  publisher={Elsevier}
}

@article{connerney2017jupiter,
  title={Jupiter’s magnetosphere and aurorae observed by the Juno spacecraft during its first polar orbits},
  author={Connerney, John EP and Adriani, Alberto and Allegrini, F and Bagenal, F and Bolton, SJ and Bonfond, Bertrand and Cowley, Stanley William Herbert and Gerard, J-C and Gladstone, GR and Grodent, Denis and others},
  journal={Science},
  volume={356},
  number={6340},
  pages={826--832},
  year={2017},
  publisher={American Association for the Advancement of Science}
}

@article{adriani2018clusters,
  title={Clusters of cyclones encircling Jupiter’s poles},
  author={Adriani, Alberto and Mura, Alessandro and Orton, G and Hansen, C and Altieri, Francesca and Moriconi, ML and Rogers, J and Eichst{\"a}dt, G and Momary, T and Ingersoll, Andrew P and others},
  journal={Nature},
  volume={555},
  number={7695},
  pages={216--219},
  year={2018},
  publisher={Nature Publishing Group UK London}
}

@inproceedings{bayer2015europa,
  title={Europa Clipper mission: the habitability of an icy moon},
  author={Bayer, Todd and Cooke, Brian and Gontijo, I and Kirby, Karen},
  booktitle={2015 IEEE aerospace conference},
  pages={1--12},
  year={2015},
  organization={IEEE}
}

@article{grasset2013jupiter,
  title={JUpiter ICy moons Explorer (JUICE): An ESA mission to orbit Ganymede and to characterise the Jupiter system},
  author={Grasset, Olivier and Dougherty, MK and Coustenis, A and Bunce, EJ and Erd, C and Titov, D and Blanc, M and Coates, A and Drossart, P and Fletcher, LN and others},
  journal={Planetary and Space Science},
  volume={78},
  pages={1--21},
  year={2013},
  publisher={Elsevier}
}

@techreport{witasse2021juice,
  title={JUICE (Jupiter Icy Moon Explorer): Plans for the cruise phase},
  author={Witasse, Olivier and Altobelli, Nicolas and Andres, Rafael and Atzei, Alessandro and Boutonnet, Arnaud and Budnik, Frank and Dietz, Angela and Erd, Christian and Evill, Ry and Lorente, Rosario and others},
  year={2021},
  institution={Copernicus Meetings}
}

@inproceedings{dougherty2012jupiter,
  title={JUpiter ICy moons Explorer (JUICE): The ESA L1 mission to the Jupiter system},
  author={Dougherty, Michele K and Grasset, Olivier and Erd, C and Titov, Dmitry V and Bunce, E and Coustenis, Ath{\'e}na and Blanc, Michel and Coates, AJ and Drossart, Pierre and Fletcher, Lyndsay and others},
  booktitle={43rd Lunar and Planetary Science Conference},
  volume={1683},
  pages={1039},
  year={2012}
}

@article{grasset2013review,
  title={Review of exchange processes on Ganymede in view of its planetary protection categorization},
  author={Grasset, Olivier and Bunce, EJ and Coustenis, Ath{\'e}na and Dougherty, Michele K and Erd, C and Hussmann, Hauke and Jaumann, Ralf and Prieto-Ballesteros, Olga},
  journal={Astrobiology},
  volume={13},
  number={10},
  pages={991--1004},
  year={2013},
  publisher={Mary Ann Liebert, Inc. 140 Huguenot Street, 3rd Floor New Rochelle, NY 10801 USA}
}

@article{nasa2024search,
  title={NASA's Europa lander and the search for life on a distant Jupiter moon},
  author={NASA},
  journal={Space Explored},
  year={2024},
  url={https://spaceexplored.com/2024/04/25/nasas-europa-lander-and-the-search-for-life-on-a-distant-jupiter-moon/}
}

@article{wall2024maps,
  title={NASA Maps Goals for Potential Landing On Jupiter's Moon Europa},
  author={Wall, Mike},
  journal={Space},
  year={2024},
  url={https://www.space.com/22291-jupiter-moon-europa-lander-mission.html}
}

@book{greenberg2005europa,
  title={Europa-the ocean moon: search for an alien biosphere},
  author={Greenberg, Richard and others},
  year={2005},
  publisher={Springer}
}

@misc{Tsai_NASA24,
    author={Yu-Dai Tsai},
title="{2024 NASA Fundamental Physics Workshop}",
      year={May 15, 2024}
}

\end{document}